\documentstyle[prc,preprint,aps,epsf]{revtex}
\newcommand{\be}{\begin{equation}}
\newcommand{\ee}{\end{equation}}
\newcommand{\ba}{\begin{eqnarray}}
\newcommand{\ea}{\end{eqnarray}}

\newcommand{\ave}[1]{\langle {#1} \rangle}
\tightenlines
\begin{document}

\title { $\Phi$-measure of pions in high energy heavy-ion collisions}
\author{Q. H. Zhang}
\address{Institute of Theoretical Science 
and Department of Physics,
University of Oregon,
Eugene, Orgeon 97403-5203\\
Physics Department, McGill University, Montreal QC H3A 2T8, Canada}
\vfill
\maketitle

\begin{abstract}
Event-by-Event fluctuations for high energy heavy-ion 
collisions are investigated. The $\Phi$-measures are calculated 
for pions. It is found that multiparticle symmetrization 
correlation has great influence on 
the value of $\Phi$ if the phase space density is high. 
Previous derivation of $\Phi$-measure has neglected 
multiparticle quantum correlation. So the explanation of 
$\Phi$-measure as a measure of dynamical correlation is 
reasonable only when multiparticles quantum correlation 
effects are smaller.  We argue that the $\Phi$-measure might 
be applicable at the SPS energy where the pion phase space density 
is much less than one.
\end{abstract}
\vspace{0.3cm}

\section{Introduction}

One of the main goals of relativistic heavy-ion physics 
is the discovery of the new state of strongly interacting 
matter that is called the quark-gluon plasma(QGP)\cite{Hwa}. 
As one can not detect QGP directly, lots of 
indirect signals are suggested\cite{Hwa}.
It has been pointed out that the study of event-by-event 
fluctuations in high energy heavy-ion collisions may provide us more 
information about the experiment\cite{CH,HZY01,stod,Tra99}. 
For examples, event-by-event fluctuations 
may provide us information about the heat capacity 
\cite{stod,blei1,shur,SRS99}, possible equilibration of the system
\cite{gazd1,gazd2,mrow1,mrow2,Liu,VKR,blei2,gazd3,Mro99}
or about the phase transition\cite{BH99,SRS99}. Experimental group NA49 
has already studied the $\Phi$-measure 
in Pb+Pb collisions at SPS energy\cite{roland}. 
Recently Bialas, Koch\cite{BK99} and 
Belkacem et. al.\cite{Belk99} have found a relationship between 
the moments of event-by-event fluctuations and the inclusive 
correlations. 
 In this paper we will extend the derivation given in\cite{Belk99}, 
and connect the result of Ref.\cite{BK99} with Ref.\cite{Belk99}.  
It has been suggested in Ref.\cite{gazd1} that if there is no 
special dynamics involved in nuclei and nuclei $(AA)$ collisions, 
then the $\Phi$-measure for $AA$ should be the same 
as the $\Phi$-measure for proton and proton $(pp)$ collisions.
So by comparing 
$\Phi$-measures for pp and AA, one can get 
information on the dynamics that is involved in 
$AA$ collisions. However, the derivation in Ref.\cite{gazd1} has neglected 
the multi-particles Bose-Einstein symmetrization 
effects among the identical particles.  The 
difference between $\Phi$-measure for $AA$ and 
that for $pp$ collisions may be due to the multi-particles 
Bose-Einstein correlation so that  
the statement of  Ref.\cite{gazd1} given above 
 is no longer valid anymore. In Ref.\cite{mrow2}, the author has assumed that 
the pions produced from $AA$ collisions has a Bose-Einstein 
distribution form in momentum space and has used this spectrum distribution 
to calculate $\Phi$-measure and found that $\Phi$-measure of 
Bose-Einstein symmetrization is around the prediction of the 
experiment. But it has been shown in Ref.\cite{ZP00} that only when 
the volume goes to infinity, the spectrum distribution of pions 
can be of the function form of Bose-Einstein. In this 
paper, we will
calculate the $\Phi$-measure of pions in heavy-ion collisions by using 
 a pion distribution function in both coordinate and momentum 
space.  As has been shown in Ref.\cite{multi} that when coordinate 
space volume goes to infinity, multi-particle Bose-Einstein 
correlation will make the pion spectrum distribution change to 
 Bose-Einstein form in momentum space. So the calculation in Ref.\cite{mrow2} 
can be achieved in our calculation by letting the volume goes to infinity.
On the other side, the effects of the finite  size of $AA$ system on  
$\Phi$-measure can be studied in our calculation. 
This paper is arranged in the following way: In section II, the relationship 
between $\Phi$-measure and particles inclusive distributions are derived.
In section III, we first derive  the formulas which will be used for the calculation 
of $\Phi$-measure of pions, then we calculate 
the $\Phi$-measure for different system size and temperature. 
For comparison, the 
$\Phi$-measure of pions 
based on the method given in Ref.\cite{mrow2} are also calculated. 
In section IV, a simple model is given which demonstrates explicitly that the 
derivation given in Ref.\cite{Mro99} has neglected multi-particles 
Bose-Einstein symmetrization effects and conclusions given in Ref.\cite{Mro99}
only holds when the multi-particles symmetrization effects are smaller.
Finally, Conclusions are given in Section V.

\section{The relationship between  $\Phi$-measure and particles inclusive distribution}

The quantity $Z(y)$ for an arbitrary event is defined as\cite{gazd1,BK99}:
\be
Z(y) = \sum_{i=1}^{N} y({\bf p_i}),
\label{eq1}
\ee
here $N$ is the multiplicity of the event and $y({\bf p_i})$
is the function of momentum ${\bf p_i}$ in the event
{\footnote{If $y({\bf p_i})$ is total momentum,  
$y({\bf p_i})=|p_i|$. If $y({\bf p_i})$ is transverse momentum,
$y({\bf p_i})=\sqrt{p_1^2+p_2^2}$. 
If $y({\bf p_i})$ is the rapidity, then we 
need to know the energy of particles $i$. If $y({\bf p_i})$ is 
electric charge of the particles, we also need to know the 
electrical charge of particles $i$. So Eq.(\ref{eq1}) 
can be generalized to 
\begin{equation}
Z(y)=\sum_{i=1}^{N} y({\bf p_i},s_i),
\label{eqff1}
\end{equation}
Here $s_i$ represents other property of particles $i$, like mass and charge. 
One can simply write Eq.(\ref{eqff1}) as
\begin{equation}
Z(y)=\sum_{i=1}^{N} y_i
\end{equation}
Here $y_i$ is the function of $i$-th particle.}}

Then the k-th order moments of $Z$ can be written down as\cite{BK99,Belk99}
\be
\ave{Z(y)^{k}} = \frac{1}{M} \sum_{j=1}^{M}\left[\sum_{i=1}^{N_j}
y({\bf p_i})\right]^{k},
\label{eq2}
\ee
where $M$ is the total number of events and $N_j$ is the multiplicity of 
the $j$-th event. 

In the following we will use  
$N$-body distribution function  
$f(N,{\bf p_1},\cdot\cdot\cdot,{\bf p_N})$ 
which is normalized as
\be
\int d{\bf p_1} \cdot\cdot\cdot d{\bf p_N} ~
f(N,{\bf p_1},\cdot\cdot\cdot,{\bf p_N}) = P(N).
\label{eq3}
\ee
Here $P(N)$ is the normalized multiplicity distribution.

Following Ref.\cite{Belk99}, we define the 
reduced s-body distribution functions for a system of
indistinguishable $N-$particles as:

\be
f_s(N,{\bf p_1}, ...,{\bf p_s}) =\cases{ \frac{N!}{(N-s)!} 
\int d{\bf p_{s+1}} \cdot\cdot\cdot d{\bf p_N} ~
f(N,{\bf p_1},\cdot\cdot\cdot,{\bf p_N})&  $s < N$\cr
N!
f(N,{\bf p_1},\cdot\cdot\cdot,{\bf p_N})&  $s = N$\cr
0& $s > N$.\cr}
\label{eq5}
\ee
But we should point out that this definition is different from that 
in Ref.\cite{Belk99}{\footnote{In Ref.\cite{Belk99},the authors 
have defined a function $f_s(N,{\bf p_1},...,{\bf p_s})$ as 
\begin{equation}
f_s(N,{\bf p_1},...,{\bf p_s})=\frac{N!}{(N-s)!}
\int d{\bf p_{s+1}}...d{\bf p_N}f_N(N,{\bf p_1},...{\bf p_N}).
\label{eec1}
\end{equation}
It is clear that when $s=N$, the above formula is "bad" 
defined while this situation does not 
exist in our definition (see Eq.(\ref{eq5})).
On the other hand, one can derive from Eq.(\ref{eq5}) 
of this paper that 
\begin{equation}
f_s(N,{\bf p_1},...,{\bf p_s})=\frac{1}{(N-s)!}
\int d{\bf p_{s+1}}...d{\bf p_N}f_N(N,{\bf p_1},...{\bf p_N}).
\label{eec2}
\end{equation}
From Eq.(\ref{eec1}) and Eq.(\ref{eec2}), 
we know that this two definitions (Ref.\cite{Belk99} and here) are different.}}

According to Eq.(\ref{eq2}), One can write out the $k$-order 
moments of $Z$ as\cite{BK99,Belk99}:

\be
\ave{Z(y)^k} = \sum_{N=0}^{\infty} \int d{\bf p_1} \cdot\cdot\cdot d{\bf p_N} ~ 
\left[\sum_{i=1}^{N} y({\bf p_i})\right]^{k} 
f(N,{\bf p_1},\cdot\cdot\cdot,{\bf p_N}).
\label{eq7}
\ee
To be clear, the first to fifth order moments of Z can be expressed as:

\be
\ave{Z(y)} = \sum_{N=0}^{\infty} \int d{\bf p_1} ~ 
y({\bf p_1}) ~ f_1(N,{\bf p_1}),
\label{eq8}
\ee
\be
\ave{Z(y)^2} = \sum_{N=0}^{\infty} \left[ \int d{\bf p_1}~  y^2({\bf p_1}) ~ 
f_1(N,{\bf p_1}) + \int d{\bf p_1}d{\bf p_2} ~ 
y({\bf p_1})y({\bf p_2}) ~ f_2(N,{\bf p_1},{\bf p_2}) \right],
\label{eq9}
\ee
\ba
\nonumber
&&\ave{Z(y)^3} = \sum_{N=0}^{\infty} \left[ \int d{\bf p_1} ~ y^3({\bf p_1}) ~ 
f_1(N,{\bf p_1}) + 3\int d{\bf p_1}d{\bf p_2} ~ 
y^2({\bf p_1})y({\bf p_2}) ~ f_2(N,{\bf p_1},{\bf p_2}) \right.\\
&&~~~~~~~~~~~~\left. + \int d{\bf p_1}d{\bf p_2}d{\bf p_3} ~ 
y({\bf p_1})y({\bf p_2})y({\bf p_3}) ~ 
f_3(N,{\bf p_1},{\bf p_2},{\bf p_3}) \right],
\label{eq10}
\ea
\begin{eqnarray}
\nonumber
&&\ave{Z(y)^4} = \sum_{N=0}^{\infty} \left[ \int d{\bf p}_1 ~ y^4({\bf p}_1) ~ 
f_1(N,{\bf p}_1) + 4\int d{\bf p}_1d{\bf p}_2 ~ 
y^3({\bf p}_1)y({\bf p}_2) ~ f_2(N,{\bf p}_1,{\bf p}_2) \right.
\nonumber\\
&&~~~~~~~~~~~~\left.
+3\int d{\bf p}_1 d{\bf p}_2 ~y^2({\bf p}_1)y^2({\bf p}_2)
f_2(N,{\bf p}_1,{\bf p}_2)\right. 
\nonumber\\
&&~~~~~~~~~~~~\left.+6 \int d{\bf p}_1d{\bf p}_2 d{\bf p}_3
y^2({\bf p_1})y({\bf p_2})y({\bf p_3}) 
f_3(N,{\bf p}_1,{\bf p}_2,{\bf p}_3) \right.\nonumber\\
&&~~~~~~~~~~~~\left. + \int d{\bf p}_1d{\bf p}_2d{\bf p}_3d{\bf p}_4 ~ 
y({\bf p}_1)y({\bf p}_2)y({\bf p}_3)y({\bf p}_4) ~ 
f_4(N,{\bf p_1,p_2,p_3,p_4}) \right],
\label{eq11}
\end{eqnarray}
\begin{eqnarray}
\nonumber
&&\ave{Z(y)^5} = \sum_{N=0}^{\infty} \left[ \int d{\bf p}_1 ~ y^5({\bf p}_1) ~ 
f_1(N,{\bf p}_1) + 5\int d{\bf p}_1d{\bf p}_2 ~ 
y^4({\bf p}_1)y({\bf p_2}) ~ f_2(N,{\bf p_1,p_2}) \right.\nonumber\\
&&~~~~~~~~~~~~
+10\int d{\bf p}_1 d{\bf{p}_2} ~y^3({\bf p}_1)y^2({\bf p}_2)
f_2(N,{\bf p}_1,{\bf p}_2)\nonumber\\
&&~~~~~~~~~~~~
 +10 \int d{\bf p}_1d{\bf p}_2 d{\bf p}_3 y^3({\bf p}_1) y({\bf p}_2)
y({\bf p}_3)f_3(N,{\bf p}_1,{\bf p}_2,{\bf p}_3)\nonumber\\
&&~~~~~~~~~~~~
 +15 \int d{\bf p}_1d{\bf p}_2 d{\bf p}_3 y^2({\bf p}_1) y^2({\bf p}_2)
y({\bf p}_3)f_3(N,{\bf p}_1,{\bf p}_2,{\bf p}_3)\nonumber\\
&&~~~~~~~~~~~~
 +10 \int d{\bf p}_1d{\bf p}_2 d{\bf p}_3 y^2({\bf p}_1) y({\bf p}_2)
y({\bf p}_3)y({\bf p}_4)f_4(N,{\bf p}_1,{\bf p}_2,{\bf p}_3,{\bf p}_4)
\nonumber\\
&&~~~~~~~~~~~~\left. + \int d{\bf p_1}d{\bf p_2}d{\bf p_3}d{\bf p}_4 d{\bf p_5} ~ 
y({\bf p_1})y({\bf p_2})y({\bf p_3})y({\bf p}_4)y({\bf p_5}) ~ 
f_5(N,{\bf p_1,p_2,p_3},{\bf p}_4,{\bf p_5}) \right].
\label{eq12}
\end{eqnarray}

\noindent
The $s$ particles inclusive distribution can be 
written down as
\begin{eqnarray}
\rho({\bf p_1,p_2,...,p_s})=\sum_{N=0}^{\infty}f_s(N,{\bf p_1,p_2,...,p_s})
\label{eq13}
\end{eqnarray}
with 
\begin{equation}
\int d{\bf p_1}...d{\bf p_s}
\rho({\bf p_1,...,p_s})=\sum_{N=0}^{\infty}P(N)\frac{N!}{(N-s)!}=\langle N(N-1)...(N-s+1)\rangle .
\label{eq14}
\end{equation}
Then we have\cite{BK99}:
\begin{eqnarray}
\ave{Z(y)}=\int y({\bf p})\rho({\bf p})d{\bf p},
\label{eqxxc}
\end{eqnarray}
\begin{eqnarray}
\ave{Z(y)^2}=
\int y^2({\bf p})\rho({\bf p})d{\bf p}+
\int d{\bf p_1}d{\bf p_2}y({\bf p_1})y({\bf p_2})\rho({\bf p_1,p_2}),
\label{eq35}
\end{eqnarray}
\begin{eqnarray}
&&\ave{Z(y)^3}=\int y^3({\bf p})\rho({\bf p})d{\bf p}
\nonumber\\
&&+3\int d{\bf p_1}d{\bf p_2}y^2({\bf p_1})y({\bf p_2})\rho({\bf p_1,p_2})
\nonumber\\
&&+\int d{\bf p_1}d{\bf p_2}d{\bf p_3}y({\bf p_1})y({\bf p_2})y({\bf p_3})
\rho({\bf p_1,p_2,p_3}),
\label{eq36}
\end{eqnarray}
\begin{eqnarray}
&&\ave{Z(y)^4}=\int y^4({\bf p})\rho({\bf p})d{\bf p}+4\int d{\bf p_1}d{\bf p_2}
y^3({\bf p_1})y({\bf p_2})
\nonumber\\
&&+3\int d{\bf p_1}d{\bf p_2}y^2({\bf p_1})y^2({\bf p_2})\rho({\bf p_1,p_2})
\nonumber\\
&&+6\int d{\bf p_1}d{\bf p_2}d{\bf p_3}y^2({\bf p_1})y({\bf p_2})
y({\bf p_3})\rho({\bf p_1,p_2,p_3})
\nonumber\\
&&+\int d{\bf p_1}d{\bf p_2}d{\bf p_3}d{\bf p_4}
y({\bf p_1})y({\bf p_2})y({\bf p_3})y({\bf p_4})
\rho({\bf p_1,p_2,p_3,p_4}),
\label{eq37}
\end{eqnarray}
and so on. One can use Eq.(\ref{eq2}) or Eqs(\ref{eq35},\ref{eq36},\ref{eq37})
 to calculate $\langle Z(y)^{K}\rangle $ separately to examine 
the consistency between those equations. The discussions about the  
consistency between those equations can be found in Ref.\cite{BK99}. 

\section{$\Phi$-measure of pions with multi-particles Bose-Einstein symmetrization 
correlations}

In the following we will only consider pions($\pi^+$ or $\pi^-$), the abundant particles produced in 
high energy collisions. 
The $N$ pions distribution function can be written down as\cite{multi}
\begin{eqnarray}
f(N,{\bf p_1,...,p_N})=P(N)\frac{P_N({\bf p_1,p_2,...,p_N})}{\int 
d{\bf p_1}...d{\bf p_N}P_N({\bf p_1,...,p_N})}
\label{eq17}
\end{eqnarray}
with
\begin{eqnarray}
P_N({\bf p_1,...,p_N})=\sum_{\sigma}\rho^{(N)}({\bf p_1},{\bf p_{\sigma(1)}})
...\rho^{(N)}({\bf p_N},{\bf p_{\sigma(N)}}).
\label{eq18}
\end{eqnarray}
Here $\rho^{(N)}({\bf p_i,p_j})$ is the Fourier transformation of the 
pion source distribution function $g^{(N)}({\bf x},{\bf K})$, 
and $g^{(N)}({\bf x},{\bf K})$ is the probability of finding a pion at point ${\bf x}$ 
with momentum ${\bf K}$. 
The superscript $N$ denotes 
those quantities are calculated for a system with pion multiplicity $N$. 
$\sigma(i)$ denotes the $i$-th element of a permutations of the 
sequence $\{ 1,2,\cdot\cdot\cdot,s\}$, and the sum over $\sigma$ denotes 
the sum over all $s!$ permutations of this sequence. 

Using Eqs.(\ref{eq5},\ref{eq17},\ref{eq18}), we have\cite{Z99,Z1}: 
\begin{eqnarray}
&&f_s(N,{\bf p_1,\cdot \cdot \cdot p_s})=\frac{P(N)}{\omega^{(N)}(N)}
\sum_{i=s}^{N}\sum_{m_1=1}^{i-(s-1)}
\sum_{m_2=1}^{i-m_1-(s-2)}\cdot\cdot\cdot 
\nonumber\\
&&~~~~~~~~\sum_{m_{s-1}=1}^{i-m_1-m_2\cdot\cdot\cdot-m_{s-2}-1}
\sum_{\sigma}
G_{m_1}^{(N)}({\bf p}_1,{\bf p}_{\sigma(1)})
G_{m_2}^{(N)}({\bf p}_2,{\bf p}_{\sigma(2)})
\nonumber\\
&&~~~~~~~~
\cdot \cdot\cdot
G_{m_{s-1}}^{(N)}({\bf p}_{s-1},{\bf p}_{\sigma(s-1)})
G_{i-m_1\cdot\cdot\cdot-m_{s-1}}^{(N)}({\bf p}_{s},{\bf p}_{\sigma(s)})
\omega^{(N)}(N-i)
\label{eq19}
\end{eqnarray}
with
\begin{eqnarray}
&&\omega^{(N)}(N)=\frac{1}{N!}
\int d{\bf p_1}...d{\bf p_N}P_N({\bf p_1,...,p_N})
\nonumber\\
&&=\frac{1}{N}\sum_{i=1}^{N}\omega^{(N)}(N-i)\int d{\bf p}G_i^{(N)}({\bf p},{\bf p}),
\nonumber\\
&&~~~\omega^{(N)}(0)=1, G_1^{N}=\rho^{N}({\bf p,p})
\label{eq20}
\end{eqnarray} 
and
\begin{equation}
G_i^{(N)}({\bf p,q})=\int \rho^{(N)}({\bf p},{\bf p_1})
d{\bf p_1}\rho^{(N)}({\bf p_1},{\bf p_2})\cdot\cdot\cdot d{\bf p_{i-1}}\rho^{(N)}({\bf p_{i-1}},{\bf q}).
\label{eq21}
\end{equation}
The superscript $N$ of $\omega^{(N)}$ and 
$G_i^{(N)}({\bf p,q})$ denote $\omega^{(N)}$ and 
$G_i^{(N)}({\bf p,q})$ are calculated using the quantities
$\rho^{(N)}({\bf p_i,p_j})$. 

The $s$-pion inclusive distribution can be expressed as
\begin{eqnarray}
&&\rho({\bf p_1,\cdot\cdot\cdot,p_s})=\sum_{N=0}^{\infty} f_s(N,{\bf p_1,...,p_s})
\nonumber\\
&&=
\sum_{N=s}^{\infty}
\sum_{i=s}^{N}
\sum_{m_1=1}^{i-(s-1)}
\sum_{m_2=1}^{i-m_1-(s-2)}
\cdot\cdot\cdot 
\sum_{m_{s-1}=1}^{i-m_1-m_2\cdot\cdot\cdot-m_{s-2}-1}
P(N)\frac{\omega^{N}(N-i)}{\omega^{N}(N)} 
\sum_{\sigma}
\nonumber\\
&&
G_{m_1}^{(N)}({\bf p}_1,{\bf p}_{\sigma(1)})
G_{m_2}^{(N)}({\bf p}_2,{\bf p}_{\sigma(2)})
\cdot \cdot\cdot
G_{m_{s-1}}^{(N)}({\bf p}_{s-1},{\bf p}_{\sigma(s-1)})
G_{i-m_1\cdot\cdot\cdot-m_{s-1}}^{(N)}({\bf p}_{s},{\bf p}_{\sigma(s)})
\nonumber\\
&&=
\sum_{m_1=1}^{\infty}\cdot\cdot\cdot 
\sum_{m_s=1}^{\infty}
\sum_{N=m_1+m_2+...+m_s}^{\infty}
P(N)\frac{\omega^{(N)}(N-m_1-m_2-...-m_s)}{\omega^{(N)}(N)} 
\nonumber\\
&&
\sum_{\sigma}
G_{m_1}^{(N)}({\bf p}_1,{\bf p}_{\sigma(1)})
\cdot \cdot\cdot
G_{m_{s}}^{(N)}({\bf p}_{s},{\bf p}_{\sigma(s)}).
\label{eq22}
\end{eqnarray}

Neglecting symmetrization effects, that is $\rho^{(N)}(p_i,p_j)=0$,  we have 
\begin{equation}
\omega^{(N)}(N)=\frac{1}{N!}
\label{eq23}
\end{equation}
and
\begin{equation}
G_{k}^{(N)}({\bf p_i,p_j})=0, ~~~~ k\ne 1 \cup {\bf p_i}\ne {\bf p_j}.
\label{eq24}
\end{equation}
Thus Eq.(\ref{eq19}) reads
\begin{eqnarray}
f_s(N,{\bf p_1,...,p_s})&=&\frac{P(N)}{\omega^{(N)}(N)}
G_1^N({\bf p_1,p_1})...G_1^N({\bf p_s,p_s})\omega^{(N)}(N-s)
\nonumber\\
&=&P(N)\frac{N!}{(N-s)!}
\rho^{N}({\bf p_1,p_1})...\rho^{N}({\bf p_s,p_s}).
\label{eq25}
\end{eqnarray}
From Eq.(\ref{eq13}) or Eq.(\ref{eq22}) we have 
\begin{eqnarray}
\rho({\bf p_1,...,p_s})=\sum_{N=s}^{\infty}N(N-1)...(N-s+1)P(N)
\rho^{N}({\bf p_1,p_1})...\rho^{N}({\bf p_s,p_s}).
\label{eq26}
\end{eqnarray}
If we neglect multi-particle symmetrization(or antisymmetrization)
effects and the correlation between particle spectrum distribution 
and multiplicity,  we immediately come 
to the conclusion that $\rho^{N}({\bf p,p})=\rho({\bf p})/\langle N\rangle$.
Of course this conclusion is based on the fact that 
we have neglected the dynamical correlation and 
kinematic correlation.  

In calculation, one normally choose the variable $y({\bf p})$ as\cite{gazd1}
\be
y({\bf p}) = x({\bf p}) - \overline{x},~~~
\overline{x} = \int d{\bf p} x({\bf p})\frac{\rho({\bf p})}{\langle N\rangle}.
\label{eq27}
\ee
If we neglect all kinds of correlations in the system,
the first three moments of $Z$ read:
\be
\ave{Z^{k}} = \int d{\bf p} y^k({\bf p})\rho({\bf p})
~~~~~~~~~~ (1\le k \le 3).
\label{eq29}
\ee
For high-order moments, Eq.(\ref{eq29}) does not hold anymore. 
This can be easily seen from Eqs.(\ref{eq11},\ref{eq12},\ref{eq37}). This 
result is different from Ref.\cite{Belk99} where Eq.(\ref{eq29}) have been 
thought to be valid for all-order moments.
Mrowczynski has also noticed this fact in Ref.\cite{Mro99}.
So $\Phi$-measure 
\be
\Phi_k =(\frac{\ave{Z^k}}{\ave{N}})^{1/k} - (\overline{y^k})^{1/k} ~~~~~~~~~~(2\le k\le 3)
\label{eq30}
\ee
with
\be
\overline{y^k} =\frac{1}{\langle N\rangle}\int \rho({\bf p})y^{k}({\bf p})d{\bf p},
\label{eq31}
\ee
is related to the quantum  many-body correlations,
the correlation between the multiplicity and momentum distribution, and 
the dynamical correlations, the last one has  
already been discussed in Ref.\cite{gazd1}. 

As we have no interest in the first two kinds of correlations, we will  
eliminate those correlations from the $\Phi$-measure.
One may get rid of multiplicity and momentum correlation by redefining the variable 
$y({\bf p})$ as 
\begin{equation}
y^{(N)}({\bf p})=x({\bf p})-\overline{x^{(N)}}
\label{eq32}
\end{equation}
with 
\begin{equation}
\overline{x^{(N)}}=\int d{\bf p}x({\bf p})\frac{f_1(N,{\bf p})}{N}.
\label{eq33}
\end{equation}
Here $\frac{f_1(N,{\bf p})}{N}$ is the normalized 
particle spectrum distribution of pions with multiplicity $N$.
After this definition one may get rid of the possible 
correlation between the multiplicity and momentum. 
In the actual calculation, one may calculate the 
average value of $x({\bf p})$ for each event(denotes as $x^e$), then 
uses the redefined $y({\bf p})=x({\bf p})-x^e$ and Eq.(\ref{eq2}) to calculate  
$\langle Z^k\rangle$. For the $\overline{y^{k}}$, one may 
calculate it using following equation
\begin{equation}
\overline{y^{k}}=\frac{1}{M}\sum_{j=1}^{M}\sum_{i=1}^{N_j}y_j({\bf p_i})^k.
\end{equation}
Here $M$ is the total number of events and $N_j$ is the multiplicity 
of the $j$-th event. $y_j({\bf p})=x({\bf p})-x_j^e$ is the quantity  
defined for the $j$-th event, and $x_j^e$ is mean value of 
$x({\bf p})$ for the $j$-th event. Using this method, one 
can calculate 
the $\Phi$-measure. 

In the following we will calculate the quantum correlation effects 
on $\Phi$-measure. Assuming  $g^{(N)}({\bf x},{\bf K})$ as a Gaussian function 
in both coordinate space and momentum space, 
$\rho^{(N)}({\bf p_i,p_j})$ can be expressed as\cite{Z99,Z1} 
\begin{eqnarray}
\rho^{(N)}({\bf p_i,p_j})&=&
\int g({\bf x},\frac{{\bf p_i}+{\bf p_j}}{2})\exp(i({\bf p_i}-{\bf p_j})\cdot x)d{\bf x}
\nonumber\\
&=&\frac{1}{(2\pi\Delta^2)^{\frac{3}{2}}}\exp(-\frac{({\bf p_i-p_j})^2R(N)^2}{2})
\exp(-\frac{({\bf p_i+p_j})^2}{8\Delta(N)^2}).
\label{eq34}
\end{eqnarray}
Here $R(N)$ and $\Delta(N)$ are the source radius and temperature 
for the events with multiplicity $N$. Of course the source radius and 
temperature may be different for different event, even though those 
events have the same multiplicity. But at this calculation we will 
neglect this fluctuation. 
In Fig.1, $\Phi_2(p_t)$ vs. mean pion multiplicity are shown
{\footnote{
Here, we will explain how to calculate 
the $\Phi$-measure given in Fig.1. According to 
Eq.(\ref{eq30},\ref{eq31}), we have 
\begin{equation}
\Phi_2(p_t)=(\frac{\langle Z^{2}\rangle}{\langle N\rangle})^{1/2}-
(\frac{1}{\langle N\rangle}\int \rho({\bf p}) \tilde{p}_t^2 d{\bf p})^{1/2},
\label{fina}
\end{equation}
with
\begin{equation}
\tilde{p}_t=p_t-\langle p_t\rangle, 
\langle p_t\rangle=\frac{1}{\langle N\rangle}\int \rho({\bf p}) p_t~~~.
\end{equation}
Here $p_t$ is transverse momentum ($p_t=\sqrt{p_1^2+p_2^2}$), 
$\langle N\rangle$ is the mean multiplicity. According to Eq.(\ref{eq35}), we have 
\begin{equation}
\langle Z^2(p_t)\rangle=\int \tilde{p}_t^2\rho(p)d{\bf p}+
\int d{\bf p_1}d{\bf p_2}\tilde{p}_{1t}\tilde{p}_{2t}\rho({\bf p_1,p_2}).
\end{equation}
According to Eq.(\ref{eq22}), the single and two-particles inclusive 
distribution can be written down as 
\begin{equation}
\rho({\bf p})=
\sum_{m_1=1}^{\infty}\sum_{N=m_1}^{\infty}P(N)
\frac{\omega(N-m_1)}{\omega(N)}G_{m_1}({\bf p,p}),
\end{equation}
\begin{eqnarray}
\rho({\bf p_1,p_2})&=&\sum_{m_1=1}^{\infty}
\sum_{m_1=2}^{\infty}
\sum_{N=m_1+m_2}^{\infty}
P(N)\frac{\omega(N-m_1-m_2)}{\omega(N)}
\nonumber\\
&&
[G_{m_1}({\bf p_1,p_2})
G_{m_2}({\bf p_2,p_1})
+G_{m_1}({\bf p_1,p_1})
G_{m_2}({\bf p_2,p_2})].
\end{eqnarray}
In the calculation we assume that $R$ and $\Delta$ are the same for all 
multiplicity, so we neglect the superscript $N$ in  Eq.(\ref{eq22}). 
From the function $\rho^{(N)}(p_i,p_j)$ given in Eq.(\ref{eq34}), 
one can calculate $G_{m_1}(p_1,p_2)$ according to Eq.(\ref{eq21}) and 
$\omega(N)$ according to Eq.(\ref{eq20}). Choosing $P(N)$ as Poisson 
form, that is $P(N)=\exp(-\langle N\rangle)\frac{\langle N\rangle^N}{N!}$, 
we can calculate $\rho({\bf p_1,p_2})$ and $\rho({\bf p})$.  Bring all 
those formulas in eq.(\ref{fina}), we get $\Phi_2(p_t)$.}}.  
It is interesting to note 
that $\Phi$ value will increase as the pion multiplicity 
goes up.  For the same multiplicity, $\Phi$ value will 
decrease as the volume or temperature drops.  All these can be understood in 
the following way: 
$\Phi_2(p_t)$ will become bigger as the phase space density 
($\sim <N>/(R\Delta)^3$) becomes higher.  
Of course, the $\Phi$ value should also depend on the pion multiplicity 
distribution\cite{Z99,Z1,HPZ}, but this effect will be 
neglected at present calculation.  Here we would like to point 
out that many authors have noticed that Bose-Einstein symmetrization 
may affect the value of $\Phi$-measure. For example, In Ref.\cite{Belk99},
similar formulas as Eqs(\ref{eq8},\ref{eq9},\ref{eq10}) of the paper have been 
derived, but the functional form of $\rho({\bf p_1,...,p_s})$ have never been 
presented. On the other hand, Mrowczynski 
has studied Bose-Einstein correlation effects 
on the $\Phi$-measure by assuming the particle spectrum 
distribution is Bose-Einstein form which corresponds to $R$ is infinity for our  
calculation\cite{multi}. However Neither Belkacem et. al. nor Mrowczynski 
has calculated multi-body Bose-Einstein symmetrization effects on 
$\Phi-$measure as done in this paper.  For comparison the 
results of Ref.\cite{mrow2}
 are also shown in Fig.1{\footnote{Here we will explain 
how to calculate the $\Phi$-measure using the 
method given in Ref.\cite{mrow2}. According to Ref.\cite{mrow2},
we can infer that 
\begin{equation}
\rho({\bf p})=\frac{V}{(2\pi)^3}\frac{1}{\exp(\beta(E-\mu))-1}.
\end{equation}
Here $V=L^3$ is the volume, $\beta=T^{-1}$ is the inverse temperature,
$\mu$ is the chemical potential. $E=\sqrt{m^2+p^2}$ with $m$ being the 
particles mass and ${\bf p}$ its momentum.
The mean multiplicity 
$\langle N\rangle$ can be expressed as
\begin{equation}
\langle N\rangle =\int d{\bf p} \rho({\bf p}).
\end{equation} 
The mean $\langle p_t\rangle$ reads 
\begin{equation}
\langle p_t\rangle=\frac{\int \rho({\bf p})p_t d{\bf p}}{\langle N\rangle}.
\end{equation}
It is easily checked that\cite{mrow2} 
\begin{equation}
\frac{\langle Z^2(p_t)\rangle}{\langle N\rangle}
=\frac{\int d{\bf p}(p_t-\langle p_t\rangle)^2 
\frac{\exp(\beta(E-\mu))}
{(\exp(\beta(E-\mu))-1)^2}}
{\int d{\bf p} \frac{1}{\exp(E-\mu)-1}}.
\end{equation}
Bring those formulas in Eq.(\ref{fina}), we can get $\Phi_2(p_t)$.}}.  
The specific volume ($V=L^3=1000 fm^3$) and 
temperature ($T=150 MeV$) are 
chosen for the formulas of Ref.\cite{mrow2} so that $\langle r^2=x^2+y^2+z^2\rangle_{box, L=10fm}
\sim \langle r^2\rangle_{Gaussian, R=3fm}$ and 
$\langle p^2\rangle_{T=150 MeV} \sim \langle p^2\rangle_{\Delta_0=0.2 GeV}$.
It is interesting to notice that 
our results is larger than the results of Ref.\cite{mrow2} when the 
phase space ($\sim (\langle r^2\rangle \langle p^2\rangle)^{3/2}$) are the same.  
$\Phi_2(p_t)$ for Ref.\cite{mrow2} 
 will also become bigger if the phase space density becomes higher. 
Here we should point out that for $\langle N\rangle \sim 100$, which is the mean 
multiplicity of $\pi^+$ or $\pi^{-}$ measured in SPS energy, our $\Phi(p_t)$ is much larger 
than the experimental value of $4.6\pm 1.5$\cite{roland}. 
However for large volume ($R=5.3 fm $ and $\Delta =0.2 GeV$) or 
high temperature($ R= 3fm $ and $\Delta=0.355 GeV$), we find that 
$\Phi\sim 7 MeV$ for $<N>\sim 100$ which is consistent with the $\Phi$ value calculated 
by two-particles correlation ($5\pm 1.5 MeV$) for Pb+Pb at SPS\cite{roland}.
Taking $V=L^3=(16)^3fm^3$ and $T=150 MeV$, we find that $\Phi$ value of 
Ref.\cite{mrow2} is around $6 MeV$ which is consistent with our 
 results and data. This suggests that for large volume 
($ R= 5.1 fm$ or $L=16 fm$ 
which is around the HBT radius of NA49\cite{HBT}), 
high-order Bose-Einstein symmetrization effects 
are not significant for SPS energy. 
So our calculation is consistent with the results of Ref.\cite{mrow2}.

\section{The shortcoming of previous derivation of $\Phi$-measure}

In the following we show that 
 the previous derivation of the $\Phi$-measure is based on 
a model which has no quantum correlation among particles.
We assume that in AA collisions, particles are produced by 
each nucleon nucleon ($NN$) collisions and the number of collisions of NN has a 
distribution function $P(M)$. Here $M$ is the number of collisions 
in each event and the collision center has a distribution 
of $\rho({\bf x})$. The probability amplitude of finding a 
pion at point ${\bf x}$ can be written down as\cite{GKW} 
\begin{equation}
J({\bf x})=\sum_{i=1}^{M}J_i({\bf x})=\sum_{i=1}^{M}J({\bf x}-{\bf x_i})
\label{eq38}
\end{equation}
Here ${\bf x_i}$ is the 
collision center coordinate. In the last step of Eq.(\ref{eq38}) we assume 
that $J_i({\bf x})$ is only the function of ${\bf x-x_i}$. 
The probability amplitude in the momentum space can be 
expressed as\cite{GKW}
\begin{equation}
\tilde{J}({\bf p})=\int d{\bf x} J({\bf x})\exp(i{\bf p}\cdot {\bf x})=
\sum_{i=1}^{M}\int exp(i{\bf p}\cdot {\bf x}) J({\bf x}-{\bf x_i})d{\bf x}
=\sum_{i=1}^{M}\tilde{J}({\bf p})\exp(i{\bf p}\cdot {\bf x_i})
\label{eq39}
\end{equation}
with
\begin{equation}
\tilde{J}({\bf p})=\int J({\bf x})\exp(i{\bf p}\cdot {\bf x})d{\bf x}.
\label{eq40}
\end{equation}
Then the $s$-pion inclusive distribution reads
\begin{eqnarray}
&&\rho({\bf p_1,...,p_s})=\sum_{M=0}^{\infty} P(M)
\int d{\bf x_1}...d{\bf x_M} g({\bf x_1})...g({\bf x_M})
|\widetilde{J}({\bf p_1})|^2...
|\widetilde{J}({\bf p_s})|^2
\nonumber\\
&&~~~~~~~~~~~|\sum_{i=1}^{M}\exp(i{\bf p_1}\cdot {\bf x_i})|^2
...|\sum_{i=1}^{M}\exp(i{\bf p_s}\cdot {\bf x_i})|^2
\label{eq41}
\end{eqnarray}
the second line of Eq.(\ref{eq41}) comes from Eq.(\ref{eq39}) 
 which is due to quantum effect. 
So the 
distribution $P(X)$ in Ref.\cite{Mro99} does not hold anymore. However if we consider 
classical particles,  we
immediately can get an equation similar to the expression of $P(X)$ 
in Ref.\cite{Mro99}.  As 
the terms in the second line of Eq.(\ref{eq41}) change to 
\begin{eqnarray}
|\sum_{i=1}^{M}\exp(i{\bf p_j}\cdot {\bf x_i})|^2\rightarrow 
\sum_{i=1}^{M}|\exp(i{\bf p_j}\cdot {\bf x_i})|^2=M
\label{eq42}
\end{eqnarray}
we have 
\begin{equation}
\rho({\bf p_1,p_2,...,p_s})=\sum_{M=0}^{M=\infty}P(M)M^s
|\widetilde{J}({\bf p_1})|^2...
|\widetilde{J}({\bf p_s})|^2.
\label{eq43}
\end{equation} 
Then  the derivation in Ref\cite{Mro99} can be applied here. In passing, 
we would like to note that the $P(X)$ distribution in Ref.\cite{Mro99}
is the source distribution, while, $\rho({\bf p_1,...,p_s})$
in our work represents particle inclusive distribution.  For illustration, we 
will write out the mean  
multiplicity $\langle N\rangle $ as:
\begin{eqnarray}
\ave{N}&=&\int \rho({\bf p_1})d{\bf p_1}
\nonumber\\
&=&\ave{M}\int d{\bf p_1}|J({\bf p_1})|^2
+\ave{M(M-1)}
\int d{\bf p_1}|J({\bf p_1}|^2 |\widetilde{g}({\bf p_1})|^2
\label{eq44}
\end{eqnarray}
with
\begin{equation}
\widetilde{g}({\bf p_1})=\int d{\bf x} g({\bf x})\exp(i{\bf p_1}\cdot {\bf x}).
\label{eq45}
\end{equation}
Here $\int d{\bf p_1}|J({\bf p_1})|^2 = n_0$ is the 
mean number of pions that are emitted from each NN collisions.
The previous relation like
$\ave{N}=\ave{M}n_0$ does not hold anymore as 
the quantum effect has extra contribution to 
the average value of $\ave{N}$.  Of course the last term 
of Eq. (\ref{eq44}) depends on the 
distribution function of the source, $g({\bf x})$, and it will become  
 smaller if the collision region is larger. This 
can be explained in quantum mechanical context: If the source 
of AA collisions is very larger and dilute pion gas 
exists, quantum effect will 
become smaller. 

\section{Discussions and Conclusions}

It has been proved that 
$\Phi$-measure will be significantly influenced by the 
multiparticle correlations effects if the particle 
phase space density is very high.  However one may exclude the 
quantum symmetrization 
or antisymmetrization effects by cuts. One may do HBT 
analysis and determine the HBT momentum region first, then throw away the identical 
particle pairs which are affected greatly by particles symmetrization effects from each event. 
After this is done 
one may use the remaining particles to re-calculate the 
$\Phi$ measure. It is interesting to mention the following 
fact:  In Ref.\cite{FH99,BPB}, the authors have already 
calculated the average phase space density for pions at SPS energy, and 
have found that the pion phase space density is $0.2$, which is 
much less than $1$. Therefore multipion symmetrization effect may not 
be significant at SPS energy.  
For RHIC energy, it is estimated that the pion phase space 
density will become higher, so one should be careful when using 
$\Phi$-measure in RHIC energy. 

In the paper, we calculate the $\Phi$-measure for a specific 
source distribution function. There is a possibility that 
the results of $\Phi$-measure depends on the function 
form of the source.  As has been stated in the paper that 
 $\Phi$-measure also depend on the fluctuation 
of the source radius and temperature for different multiplicity.
All this kind of work has not been studied in detail in the 
paper as the main interest of present paper is to study the 
Bose-Einstein correlation effects on the $\Phi$-measure. 
We find that $\Phi$-measure is greatly affected by 
Bose-Einstein symmetrization if the source density 
of pions is high.
 
In this paper, we derive the relationship between 
$\Phi$-measure and particles inclusive distributions.
It is shown that besides dynamical correlation,
$\Phi$-measure is also related to the many-body 
quantum (anti)symmetrization 
correlations. 
Starting from a pion source distribution in coordinate and 
momentum space, we calculate the $\Phi$-measure for pions. It is 
found that $\Phi$-measure is strongly affected by 
multiparticle symmetrization effects if the phase space 
density is very high.  So the previous statement that 
$\Phi$-measure reflects the dynamical mechanism 
that exists in the AA does not hold anymore if the pion 
phase space density is high. 
For a very large system, the density of pions may be smaller, 
in this case the  
multiparticles symmetrization effect can be neglected. 
Finally we argue that 
one may more or less exclude the quantum symmetrization effect by 
cutting small momentum pairs of pions from each event.

\begin{center}
{\bf Acknowledgement}
\end{center}
The author thanks Prof. R. C. Hwa for reading the paper and Dr. S. Mrowczynski 
for helpful communication. 
This work was supported by US Department 
of Energy under Grant No. DE-FG03-96ER40972, NSERC of Canada and FCAR of the Quebec Government.

\begin{figure}[t]\epsfxsize=14cm \epsfysize=14cm
\centerline{\epsfbox{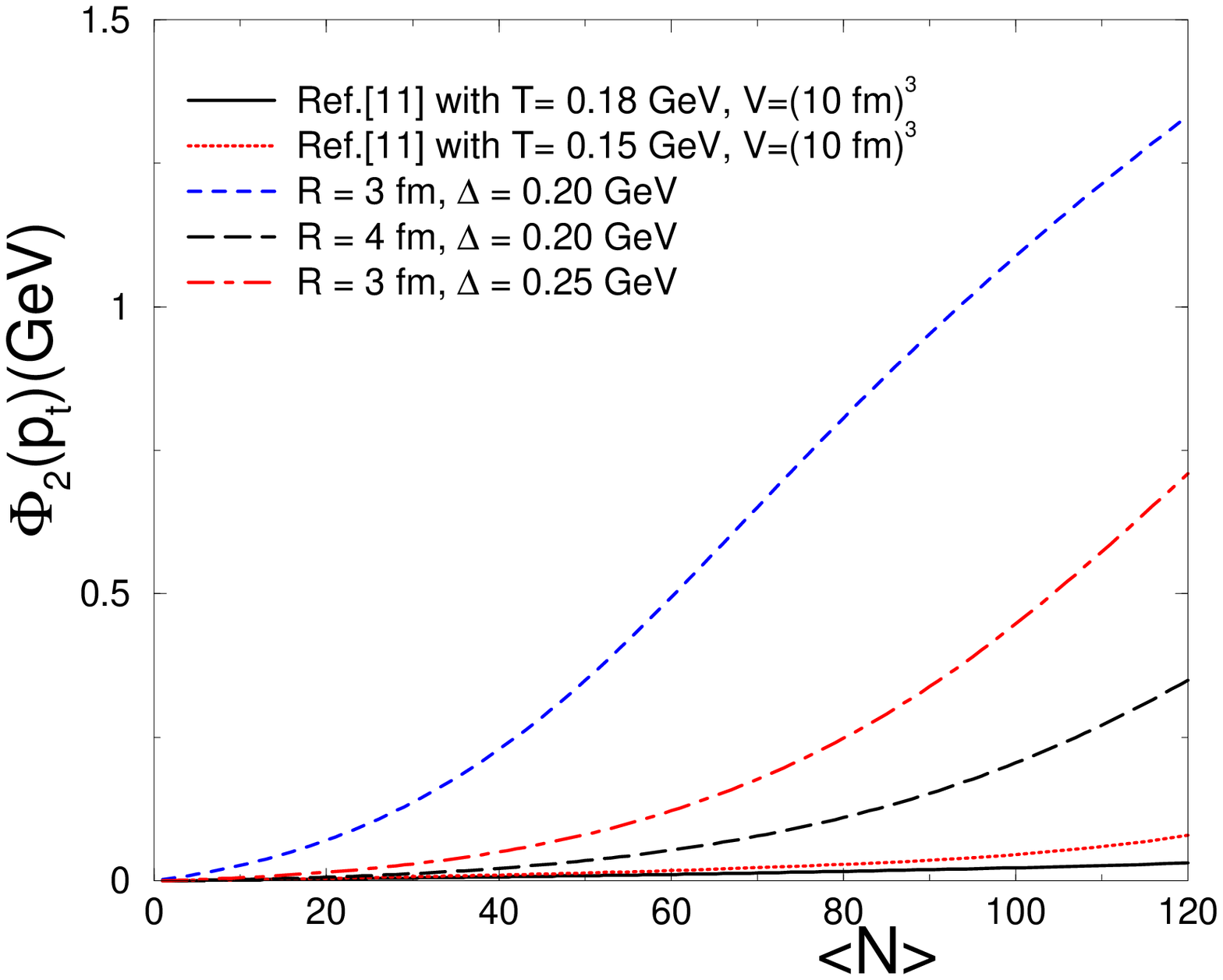}}
\caption{$\Phi_2(p_t)$ vs. mean pion multiplicity $\langle N\rangle$. In 
the calculation we choose $\Delta = 0.2 GeV$, $R = 3fm$(dashed line) 
and $4 fm$ (long dashed line) 
respectively. For comparison, $\Phi_2(p_t)$ for 
$\Delta =0. 25 GeV, R=3fm$(dot-dashed line) are also shown in Fig.1. 
We choose the pion multiplicity distribution, $P(N)$, as 
Poisson form.  The results from Ref.[11] are also 
shown in Fig.1. We choose the volume of the pion system 
(box) as $V=L^3=1000 fm^3$ and the temperature 
$T=0.15 GeV$ (dotted line) and $T=0.18 GeV$ (solid line) respectively.}
\end{figure}
\end{document}